\documentclass[11pt,a4paper]{article}

\usepackage[utf8]{inputenc}
\usepackage[T1]{fontenc}
\usepackage[british]{babel}
\usepackage{lmodern}
\usepackage{microtype}
\usepackage[a4paper,margin=2.5cm]{geometry}
\usepackage{amsmath,amssymb,amsthm}
\usepackage{bm}
\usepackage{graphicx}
\usepackage{booktabs}
\usepackage[numbers,sort&compress]{natbib}
\usepackage{hyperref}
\hypersetup{colorlinks=true,linkcolor=blue!50!black,citecolor=blue!50!black,urlcolor=blue!50!black}
\usepackage{xcolor}
\usepackage{caption}
\usepackage{authblk}
\usepackage{tikz}
\usetikzlibrary{positioning,arrows.meta,fit,backgrounds,shapes.geometric,calc}
\definecolor{cTrain}{RGB}{38,103,158}   
\definecolor{cFroze}{RGB}{110,110,116}   
\definecolor{cHead}{RGB}{30,120,110}     
\definecolor{cLoss}{RGB}{183,110,30}     
\usepackage{algorithm}
\usepackage{algpseudocode}

\newcommand{\Lam}{\Lambda}
\newcommand{\mll}{m_{\ell\ell}}
\newcommand{\sigsm}{\sigma_{\rm SM}}

\newcommand{\psib}{\bm{\psi}}

\title{\bfseries ALETHEIA: Autonomous Loop for Experimental Theory and HEP Inference Across-data}

\author[1,2]{Vincent Croft}
\affil[1]{Google Cloud Rapid Agent Hackathon}\affil[2]{Arize AX}

\begin{document}
\maketitle

\begin{abstract}
\noindent
ALETHEIA is a self-completing tool for monitoring the learning of manifolds in physics foundation models from data.
It provides a method to automatically build physics foundation models for permutation-invariant per-event representations of unknown physics manifolds. 
This process is demonstrated here for dimension-six Standard Model Effective Field Theory (SMEFT) content of four operators in neutral-current Drell-Yan, whose input is unordered event-level features, and we drive it with an active-learning loop that separates two jobs that the literature usually conflates.
\emph{Active learning completes a representation}: given a fixed operator content, an acquisition rule chooses the working points that pin the model's coefficients fastest.
\emph{The physics expands it}: which new operator to switch on is read from the residual structure, ordered by SMEFT power counting, never guessed by the acquisition.
The representation is the {\bf ManifoldInformer}, a permutation-invariant per-event
encoder $\psib_\theta$ pooled into a closed-form ridge head; its latent recovers
the analytic morphing tangents ($R^2=0.999$) and curvatures ($R^2=0.954$) of
the SMEFT cross section.
The loop monitors a residual-operator fingerprint: when a single out-of-span direction dominates, it appends that direction to $\psib_\theta$ ($\psi$-extension) and refits.
On the analytic oracle at $4000$ probe events the first $\psi$-extension collapses the per-cycle peak residual singular value from $5.36$ to $0.173$ (a factor $31$), and across the four extensions ($D_\psi: 5\!\to\!9$) $\sigma_1$ falls from $4.03$ to $0.026$ (a factor $\sim\!150$); $\sigma_1$ then stays below the completion floor and \emph{unlocking the subleading vertex operators triggers no further extension}, the operational signature that the manifold is span-complete.
The acquisition arm unlocks new operators through an Arize-Phoenix span, such that the concepts of ``learning correctly'', in which each extension collapses $\sigma_1$; and ``learned completely'', in which $\sigma_1$ is below the noise floor; are read directly off the monitored trace.
\end{abstract}

\begin{figure}[t]
\centering
\begin{tikzpicture}[
  font=\small, every node/.style={align=center},
  box/.style={rectangle, draw=black!75, rounded corners=1.5pt, minimum height=11mm,
              minimum width=30mm, line width=0.45pt, inner sep=3.5pt},
  emph/.style={box, draw=black!95, line width=0.9pt, fill=black!4},
  mon/.style={rectangle, draw=black!95, rounded corners=1.2pt, minimum height=11mm,
              minimum width=32mm, line width=0.9pt, inner sep=3pt, fill=black!8, font=\scriptsize},
  arr/.style={-{Latex[length=2.4mm,width=1.8mm]}, line width=0.55pt, draw=black!75},
  arrb/.style={-{Latex[length=2.6mm,width=2.0mm]}, line width=0.85pt, draw=black!90},
  lbl/.style={font=\scriptsize, midway, fill=white, inner sep=1.3pt},
]
\node[box] (oracle)  at (0,    2.5)  {event-level oracle\\[1pt]\scriptsize draw $N$ events at $\bm c$};
\node[box] (encoder) at (5.2,  2.5)  {encoder $\psib_\theta$\\[1pt]\scriptsize Deep Sets / EFN $\to$ pool};
\node[box] (ridge)   at (10.4, 2.5)  {ridge head\\[1pt]\scriptsize closed-form $\bm Z_{\rm pred},\ \bm A^{-1}$};
\node[mon] (mon)     at (10.4,-1.6)  {residual-SVD fingerprint\\[1pt]\scriptsize(the monitor)};
\node[box] (ctrl)    at (5.2, -1.6)  {Phoenix controller\\[1pt]\scriptsize span-completeness};
\node[box] (acq)     at (0,   -1.6)  {acquire $\bm c_{\rm cand}$\\[1pt]\scriptsize \emph{AL completes}};
\node[emph] (psi)    at (7.8,  0.45) {$\psi$-extension\\[1pt]\scriptsize \emph{physics expands}};
\draw[arr] (oracle.east)  -- node[lbl, above] {$\{x_n\}$} (encoder.west);
\draw[arr] (encoder.east) -- node[lbl, above] {rows} (ridge.west);
\draw[arr] (ridge.south)  -- node[lbl, right] {residuals $r_{\bm c}$} (mon.north);
\draw[arr] (mon.west)  -- node[lbl, above] {$\sigma_1,\,\sigma_1/\sigma_2$} (ctrl.east);
\draw[arr] (ctrl.west) -- node[lbl, above] {in-span} (acq.east);
\draw[arr] (acq.north) -- node[lbl, left] {simulate} (oracle.south);
\draw[arrb] (ctrl.north) |- node[lbl, left, pos=0.3] {out-of-span} (psi.west);
\draw[arrb] (psi.north) -- ++(0,0.55) -| node[lbl, above, pos=0.25] {append $\bm u_1$}
            ([xshift=4mm]encoder.south);
\end{tikzpicture}
\caption{\label{fig:loop} The ALETHEIA self-completing loop: the acquisition step \emph{completes} the current representation by choosing working points; the residual-SVD fingerprint detects a missing direction and the $\psi$-extension step \emph{expands} the representation by appending it; operator content is unlocked in SMEFT power-counting order, never chosen by the acquisition.}
\end{figure}
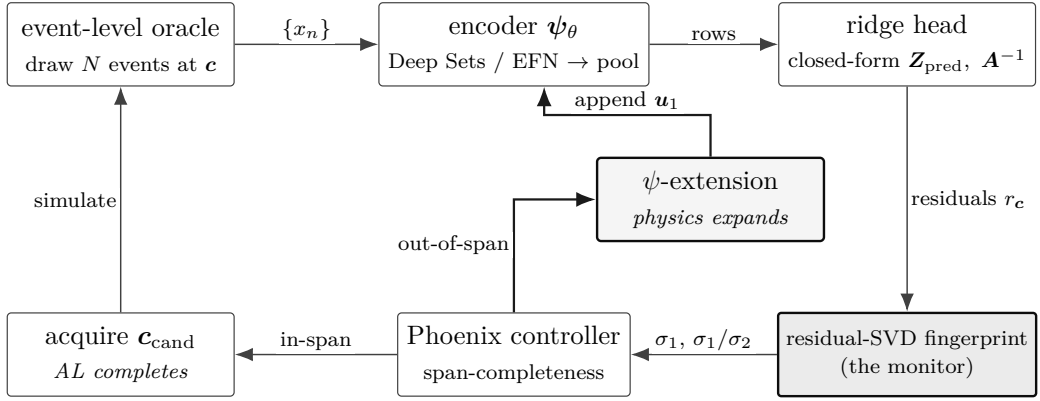

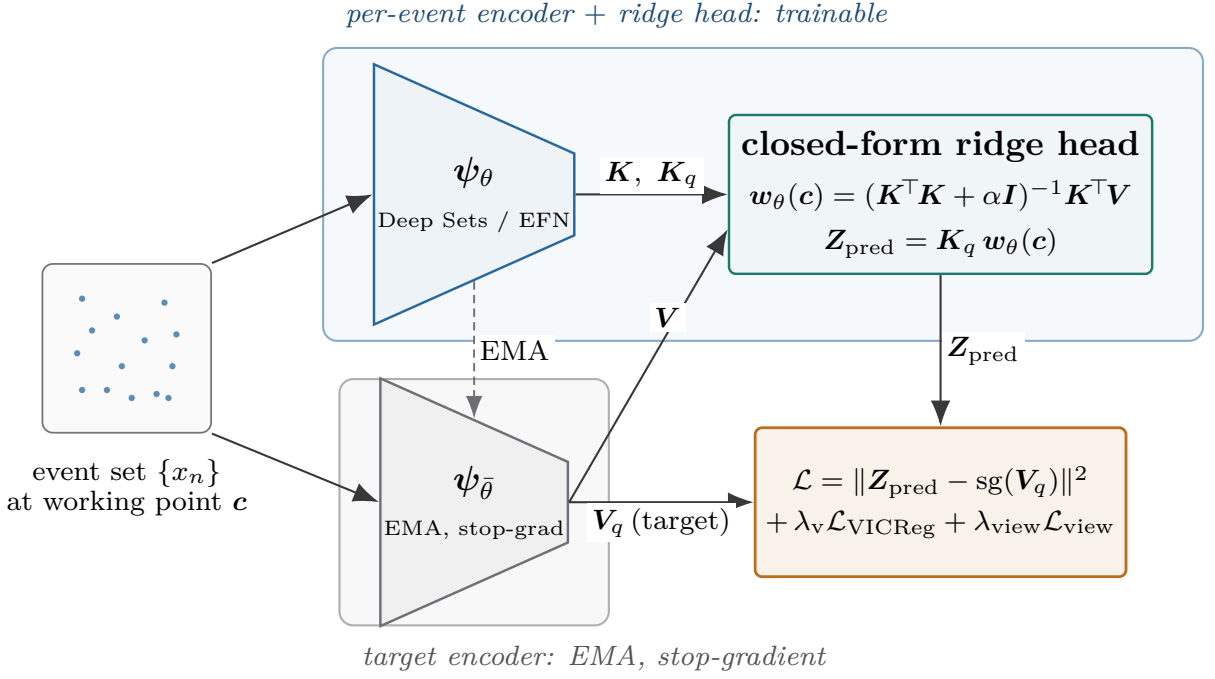
\begin{figure}[t]
\centering
\resizebox{\linewidth}{!}{%
\begin{tikzpicture}[
  font=\small, every node/.style={align=center},
  enc/.style={trapezium, shape border rotate=270, trapezium angle=66,
              draw=cTrain, fill=cTrain!8, line width=0.7pt,
              minimum width=20mm, minimum height=17mm, inner sep=1pt, text=black},
  tenc/.style={enc, draw=cFroze, fill=cFroze!10},
  head/.style={rectangle, rounded corners=2pt, draw=cHead, fill=cHead!7,
               line width=0.7pt, minimum height=16mm, minimum width=40mm, inner sep=3.5pt},
  loss/.style={rectangle, rounded corners=2pt, draw=cLoss, fill=cLoss!9,
               line width=0.8pt, minimum height=15mm, minimum width=30mm, inner sep=3pt, font=\scriptsize},
  cloud/.style={rectangle, rounded corners=3pt, draw=black!55, fill=black!2,
                line width=0.5pt, minimum width=17mm, minimum height=17mm},
  arr/.style={-{Latex[length=2.6mm,width=2.0mm]}, line width=0.6pt, draw=black!78},
  emaarr/.style={-{Latex[length=2.3mm,width=1.7mm]}, line width=0.55pt, draw=cFroze,
              dash pattern=on 2.4pt off 1.5pt},
  lbl/.style={font=\scriptsize, midway, fill=white, inner sep=1.4pt},
  panel/.style={rounded corners=4pt, line width=0.6pt},
]
\node[cloud] (inp) at (0,0) {};
\foreach \x/\y in {-0.45/0.5, -0.1/0.32, 0.38/0.46, 0.18/0.08, -0.5/-0.05,
                -0.2/-0.42, 0.46/-0.18, 0.05/-0.5, 0.5/0.14, -0.35/0.18,
                0.3/-0.46, -0.05/-0.18, 0.42/-0.5, -0.45/-0.42}
   \fill[cTrain!75] ([shift={(\x,\y)}]inp.center) circle (0.9pt);
\node[font=\scriptsize, align=center, below=1.2mm of inp]
   {event set $\{x_n\}$\\[-1pt]at working point $\bm c$};
\node[enc]  (phi) at (3.5, 1.55) {$\psib_\theta$\\[1pt]\tiny Deep Sets / EFN};
\node[tenc] (ema) at (3.5,-1.55) {$\psib_{\bar\theta}$\\[1pt]\tiny EMA, stop-grad};
\node[head] (ridge) at (8.2, 1.55)
   {\textbf{closed-form ridge head}\\[2pt]\scriptsize
    $\bm w_\theta(\bm c)=(\bm K^{\!\top}\!\bm K+\alpha\bm I)^{-1}\bm K^{\!\top}\!\bm V$\\[1pt]
    \scriptsize $\bm Z_{\rm pred}=\bm K_q\,\bm w_\theta(\bm c)$};
\node[loss] (loss) at (8.2,-1.55)
   {$\mathcal{L}=\lVert\bm Z_{\rm pred}-\mathrm{sg}(\bm V_q)\rVert^2$\\[2pt]
    $+\,\lambda_{\rm v}\mathcal{L}_{\rm VICReg}+\lambda_{\rm view}\mathcal{L}_{\rm view}$};
\draw[arr] (inp.north east) -- (phi.west);
\draw[arr] (inp.south east) -- (ema.west);
\draw[arr] (phi.east)  -- node[lbl, above] {$\bm K,\ \bm K_q$} (ridge.west);
\draw[arr] (ema.east)  -- node[lbl, below] {$\bm V_q$\,(target)} (loss.west);
\draw[arr] (ema.east)  -- node[lbl, above, pos=0.62] {$\bm V$} ([yshift=-3.5mm]ridge.west);
\draw[arr] (ridge.south) -- node[lbl, right] {$\bm Z_{\rm pred}$} (loss.north);
\draw[emaarr] (phi.south) -- node[lbl, right] {EMA} (ema.north);
\begin{scope}[on background layer]
  \node[panel, draw=cTrain!55, fill=cTrain!4,
        fit=(phi)(ridge), inner xsep=5mm, inner ysep=6mm] (ptrain) {};
  \node[panel, draw=cFroze!55, fill=cFroze!5,
        fit=(ema), inner xsep=4mm, inner ysep=4mm] (ptarg) {};
\end{scope}
\node[cTrain!72!black, font=\scriptsize\itshape, anchor=south west]
   at ([xshift=1mm,yshift=0.6mm]ptrain.north west) {per-event encoder $+$ ridge head: trainable};
\node[cFroze!75!black, font=\scriptsize\itshape, anchor=north west]
   at ([xshift=1mm,yshift=-0.6mm]ptarg.south west) {target encoder: EMA, stop-gradient};
\end{tikzpicture}}
\caption{\label{fig:arch} The ALETHEIA ManifoldInformer: a shared permutation-invariant per-event encoder $\psib_\theta$ embeds context and held-out events; the closed-form ridge head solves for the manifold coordinate $\bm w_\theta(\bm c)$ and predicts the held-out embeddings $\bm Z_{\rm pred}=\bm K_q\bm w_\theta(\bm c)$, matched in latent space against an EMA target encoder $\psib_{\bar\theta}$ with VICReg and an RS3L view-invariance term. The Wilson coefficient $\bm c$ never enters the forward pass; it indexes the working point.}
\end{figure}

\section{Introduction}
\label{sec:intro}

Physics is not that complicated. A scientist proposes a model, nature provides data, and the experiment adjusts the model to fit.
With ALETHEIA we automate this entire cycle using Agentic AI.  
A foundation model trained on data from an oracle performs inference over a family of simulator outputs using active learning to decide where to sample next.
Within a fixed operator content, active learning completes the representation by pinning the coefficients with fewer samples than random.
The direction of model expansion is read from the residual structure of the current fit, ordered by the SMEFT power counting. 
This is the physics prior on which operators enter first. 
The mathematical properties of the perturbative expansion on with SMEFT is constructed mean that it is not possible to use active learning to also guide the expansion, though this may also be possible for learning other physics theories. 
The result is a loop demonstrated in figure~\ref{fig:loop} that grows its own representation operator by operator and stops when the data shows nothing is missing.

The vehicle is neutral-current Drell-Yan with dimension-six SMEFT operators,
where the four-fermion operators $\mathcal{O}_{\ell q}^{(1,3)}$ enter the rate at
$\mathcal{O}(\hat s/\Lam^2)$ and dominate the high-mass tail, while the vertex
operators $\mathcal{O}_{Hq}^{(1,3)}$ enter at $\mathcal{O}(v^2/\Lam^2)$ as a
near-normalisation shift that the mass spectrum cannot distinguish from
luminosity. 
This is exactly a leading-then-subleading hierarchy: complete the
four-fermion manifold first, then turn on the vertex operators. The
representation is the ManifoldInformer (Section~\ref{sec:fm}); the oracle and its
operator structure are in Section~\ref{sec:oracle}; the complete-then-expand
algorithm is in Section~\ref{sec:algo}; Section~\ref{sec:results} traces the
improvement on the analytic oracle and shows how correctness and completeness are
monitored in Phoenix.

Framed as inference, the per-event log-likelihood ratio $\log w_{\bm c}(x)$ is
the object an amortised simulation-based-inference
estimator~\cite{Cranmer:2020frontiers,Brehmer:2018kdj,Brehmer:2019xox} targets.
The SMEFT morphing structure of Eq.~(\ref{eq:morphing}) makes it exact rather
than learned, and the analytic oracle is the validation harness for a method
whose deployment target is a high-fidelity simulator.
The active-learning loop is then simulation-based inference with a learned, self-completing representation.

\section{The oracle and its operator structure}
\label{sec:oracle}

\begin{figure}[t]
\centering
\includegraphics[width=\linewidth]{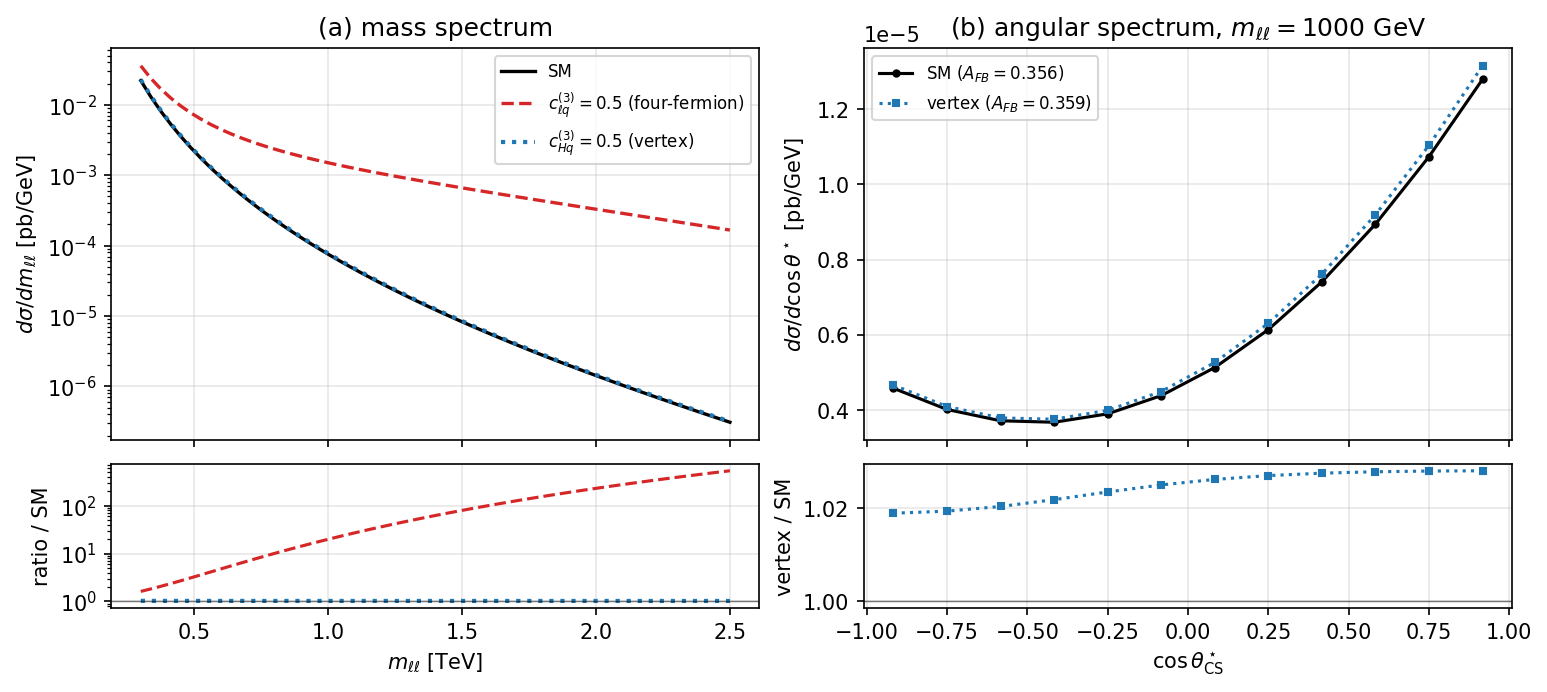}
\caption{\label{fig:phys} Oracle distributions behind Table~\ref{tab:operators},
at the working points used by the loop. (a) Mass spectrum $d\sigma/d\mll$ for the
SM, a four-fermion point ($c_{\ell q}^{(3)}=0.5$) and a vertex point
($c_{Hq}^{(3)}=0.5$), with the lower panel the ratio to SM: the four-fermion rate
grows $331\times$ into the tail, the vertex is a flat $2.5\%$ shift. (b) Angular
spectrum $d\sigma/d\cos\theta^\star$ at $\mll=1$~TeV; the vertex/SM ratio has a
slope in $\cos\theta^\star$ and shifts $A_{\rm FB}$ by $+0.0026$, so the vertex
carries forward-backward-asymmetry structure rather than a pure normalisation
change.}
\end{figure}

A \emph{working point} is a value of the Wilson-coefficient vector $\bm c$ at
which the oracle generates events. The dimension-six SMEFT Lagrangian is
$\mathcal{L} = \mathcal{L}_{\rm SM} + \sum_{i=1}^{n} (c_i/\Lam^2)\,\mathcal{O}_i$
in the Warsaw basis~\cite{Grzadkowski:2010es}, with $\Lam$ the effective-field-theory
scale, $n$ the operator count, $v$ the Higgs vacuum expectation value, and
$\hat s$ the partonic Mandelstam invariant. The squared amplitude is pointwise
quadratic in $\bm c$ in the event kinematics $x$,
\begin{equation}
  |\mathcal{M}(\bm c;x)|^2 = |\mathcal{M}_{\rm SM}(x)|^2
    + \sum_i c_i\,2\,\mathrm{Re}\!\big[\mathcal{M}_{\rm SM}^\ast(x)\,\mathcal{M}_i(x)\big]
    + \sum_{i\le j} c_i c_j\,\mathrm{Re}\!\big[\mathcal{M}_i^\ast(x)\,\mathcal{M}_j(x)\big],
  \label{eq:amp}
\end{equation}
so the differential rate is itself pointwise quadratic,
\begin{equation}
  \frac{d\sigma}{dx}(\bm c;x) = \frac{d\sigsm}{dx}(x)
    + \sum_i A_i(x)\,c_i + \sum_{i\le j} B_{ij}(x)\,c_i c_j,
  \qquad x=(\log\mll,\;\cos\theta^\star),
  \label{eq:morphing}
\end{equation}
where $A_i(x)=\partial\sigma/\partial c_i|_{\bm c=0}$ are the morphing
\emph{tangents} (the linear, interference coefficients) and $B_{ij}(x)$ the
\emph{curvatures} (the quadratic, squared-amplitude coefficients), both fixed
functions of $x$. The dependence on $\bm c$ is exact event by event, before any
binning or integration, which is what makes the morphing basis exact at event
level and not merely for the integrated rate. The per-event likelihood ratio
against the Standard Model follows,
\begin{equation}
  w_{\bm c}(x) = \frac{d\sigma/dx\,(\bm c;x)}{d\sigsm/dx\,(x)}
    = 1 + \sum_i a_i(x)\,c_i + \sum_{i\le j} b_{ij}(x)\,c_i c_j,
  \quad a_i \equiv \frac{A_i}{d\sigsm/dx},\;\;
  b_{ij}\equiv \frac{B_{ij}}{d\sigsm/dx},
  \label{eq:lr}
\end{equation}
and $\log w_{\bm c}(x)$ is the per-event log-likelihood ratio the encoder
represents and the loop residualises via $\Pi_{\psib}$ in Section~\ref{sec:algo}.
A finite basis of $N=\binom{n+2}{2}$ working points fixes the templates exactly,
so the oracle is differentiable in $\bm c$ in closed form. We use the analytic
leading-order oracle (the \texttt{analytic\_smeft} module: closed-form partonic
amplitudes convolved with CT18NNLO PDFs~\cite{Hou:2019ct18,Buckley:2014lhapdf}),
cross-validated against a MadGraph pipeline~\cite{Alwall:2014mg5}, restricted to
the four operators of Table~\ref{tab:operators}; the $(1)$ and $(3)$ superscripts
denote weak-isospin singlet and triplet contractions. Figure~\ref{fig:phys}
shows the two regimes computed from the oracle: the four-fermion tail grows by
two orders of magnitude while the vertex operator is a flat $2.5\%$ rate shift
that nonetheless tilts the angular spectrum.

\begin{table}[t]
\centering\small
\caption{\label{tab:operators} The four operators and their power counting. The
loop completes the leading tier before unlocking the subleading one. The vertex
operators are a flat $\sim\!2.5\%$ rate shift on the mass spectrum (degenerate
with normalisation); the four-fermion rate grows two orders of magnitude into the
tail. Grouping follows~\cite{Greljo:2017vvb,DasBakshi:2025reusable}.}
\begin{tabular}{cll}
\toprule
tier & power counting & operators \\
\midrule
leading     & $\mathcal{O}(\hat s/\Lam^2)$  & $\mathcal{O}_{\ell q}^{(3)},\ \mathcal{O}_{\ell q}^{(1)}$ (four-fermion) \\
subleading  & $\mathcal{O}(v^2/\Lam^2)$     & $\mathcal{O}_{Hq}^{(3)},\ \mathcal{O}_{Hq}^{(1)}$ (vertex) \\
\bottomrule
\end{tabular}
\end{table}

\section{The ManifoldInformer foundation model}
\label{sec:fm}

In this work we propose a novel neural architecture called a ManifoldInformer demonstrated in figure ~\ref{fig:arch}.
The exchangeable set that is fed to the model is the collection of events drawn at a working point.
Each event is a kinematic vector $x=(\log\mll,\cos\theta^\star)$, where $\cos\theta^\star$ is the lepton polar angle in the Collins-Soper frame; the
per-event encoder $\psib_\theta$ lifts it to $\mathbb{R}^{D_\psi}$, and a Deep
Sets / Energy-Flow pooling~\cite{Zaheer:2017deepsets,Komiske:2019efn} over the
event set, not over particles within an event, produces the working point's
context row.
Stacking context rows over working points gives the input to the closed-form ridge head~\cite{Garnelo:2023intention}: with $\bm K$ the context-event embeddings (keys), $\bm V$ their exponential-moving-average targets (values), $\alpha$ the ridge parameter and $\bm K_q$ the held-out query embeddings, it solves
$\bm w_\theta(\bm c)=(\bm K^{\!\top}\bm K+\alpha\bm I)^{-1}\bm K^{\!\top}\bm V$ and
predicts $\bm Z_{\rm pred}=\bm K_q\,\bm w_\theta(\bm c)$ in one pass; we call
$\bm w_\theta(\bm c)$ the manifold coordinate of working point $\bm c$. The head
also returns a closed-form predictive variance and a rank-one Sherman--Morrison
sequential update~\cite{Sherman:1950}, used by the EPIG~\cite{Smith:2023epig}
acquisition; the runs here use the residual-energy rule of
Section~\ref{sec:algo}. Training follows the Joint-Embedding Predictive
Architecture~\cite{LeCun:2022path,Assran:2023ijepa} with a variance-covariance
regulariser against collapse~\cite{Bardes:2022vicreg} and re-simulation view
invariance~\cite{Harris:2024rs3l}; a view is two independent Monte Carlo event
draws at the same working point $\bm c$, since with an analytic oracle there is
no shower to re-run.
In this architecture a shared per-event encoder feeds
the closed-form ridge head, whose solve returns the manifold coordinate
$\bm w_\theta(\bm c)$, and the predicted embeddings are matched against an
exponential-moving-average target encoder in the JEPA pretext.

\begin{figure}[t]
\centering
\includegraphics[width=0.92\linewidth]{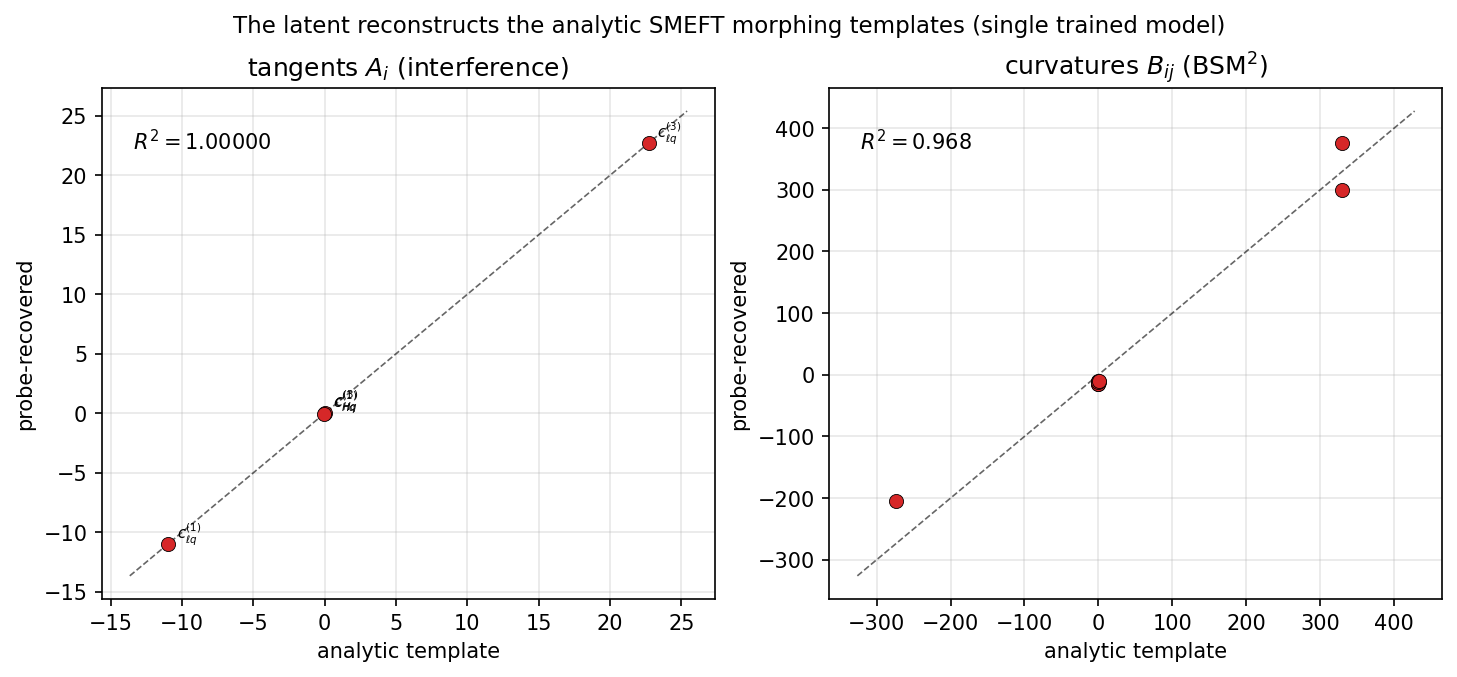}
\caption{\label{fig:recovery} The latent reconstructs the analytic SMEFT morphing
templates. (a) Probe-recovered tangents against the analytic tangents $A_i$, one
marker per operator; (b) curvatures against $B_{ij}$, one marker per operator
pair; dashed identity line. The probe regresses the encoder's Wilson-coefficient
tangent (resp.\ curvature) onto the $m$-averaged template per operator. This
single trained model attains $R^2=1.0000$ (tangents) and $R^2=0.968$
(curvatures); the headline $R^2=0.99999/0.954$ quoted in the text are medians
over $15$ retrained seeds at the same protocol.}
\end{figure}

The latent is validated against the analytic structure of
Eq.~(\ref{eq:morphing}): a linear probe from the latent onto the morphing
tangents $A_i$ recovers them at $R^2=0.99999$, and onto the curvatures $B_{ij}$
at $R^2=0.954$ (medians over $15$ retrained seeds; Figure~\ref{fig:recovery}).
The encoder therefore \emph{is} the analytic morphing geometry, learned from
events; this is the FM contract the loop builds on. Among collider foundation models (jet-level
self-supervised pretraining such as masked-particle
modelling~\cite{Heinrich:2024mpm}, OmniJet-$\alpha$~\cite{Birk:2024omnijet},
OmniLearn~\cite{Mikuni:2024omnilearn}, HEP-JEPA~\cite{Bardhan:2025hepjepa}, and
the SMEFT contrastive embedding of~\cite{DasBakshi:2025reusable}), this is the
event-level, parton-level, regression corner, and the only one whose latent is
graded by recovery of an analytic physics manifold rather than a downstream
classifier.

\newpage
\begin{quote}
\textbf{Proposition (identifiability).} The likelihood ratio $w_{\bm c}$ is the
fixed algebraic function (Eq.~\ref{eq:lr}) of the templates $\{A_i,B_{ij}\}$ and
$\bm c$, and the ridge head is linear in the encoder output. The encoder's
Wilson tangent $\partial\bm w_\theta/\partial c_i|_{\bm c=0}$ and curvature
$\partial^2\bm w_\theta/\partial c_i\partial c_j$ therefore carry the morphing
tangents $A_i$ and curvatures $B_{ij}$ as linear read-outs precisely when those
templates lie in the feature span,
$\mathrm{span}\{\psib_\theta(x)\}\supseteq\{A_i(x),B_{ij}(x)\}$ on the probe
grid. A linear probe attains $R^2\to1$ exactly under that containment, so the
measured $R^2$ (Figure~\ref{fig:recovery}) estimates it. Span-completeness
(Section~\ref{sec:algo}) is the same statement read through the residual: every
$\log w_{\bm c}$ projects into $\mathrm{span}\{\psib_\theta\}$ to within the
floor $\tau$. The recovery $R^2$ and the completion criterion $\sigma_1\le\tau$
are thus one property measured two ways.
\end{quote}

\section{The complete-then-expand algorithm}
\label{sec:algo}

The loop holds a representation $\psib_\theta$ and grows it. A working point
$\bm{c}$ contributes a residual column $r_{\bm c}(x_n) = \log w_{\bm c}(x_n) -
\Pi_{\psib}\log w_{\bm c}(x_n)$, the part of its event-level log-likelihood ratio
the current encoder cannot represent, evaluated on the $E=4000$ probe events
($\Pi_{\psib}$ projects onto the span of $\psib_\theta$). The residual operator
$\bm{R}$ is the $E\times|H|$ matrix whose columns are these raw, unnormalised
per-event residual vectors, one per acquired working point in the history $H$;
its leading singular value $\sigma_1$ therefore scales as $\sqrt{E}$ times a
typical per-event residual, and measures how much out-of-span structure remains,
while the ratio $\sigma_1/\sigma_2$ is the fingerprint of a \emph{single} missing
direction. Two actions follow. \textbf{Acquisition} (AL completes) is the
residual-energy rule: it picks the next working
point maximising $\|r_{\bm c}\|\,|\bm{u}_1^\top r_{\bm c}|$, the out-of-span
energy aligned with the current leading residual direction, exposing missing
structure fastest. EPIG~\cite{Smith:2023epig} is the closed-form target-oriented
alternative; the runs here use this residual-energy rule.
\textbf{$\psi$-extension} (physics expands) appends the dominant residual
direction $\bm{u}_1$ to $\psib_\theta$ when $\sigma_1$ exceeds the completion
floor $\tau$. The floor $\tau=0.30$ is a fixed hyperparameter on the $\sigma_1$
scale just defined ($\sqrt{E}$ times a per-event residual); no sensitivity sweep
is claimed. Completing the leading manifold (acquisition no longer raises
$\sigma_1$) unlocks the next power-counting tier.

\begin{algorithm}[t]
\caption{Complete-then-expand span-completeness loop}
\label{alg:loop}
\begin{algorithmic}[1]
\Require oracle $O$, probe events $\{x_n\}$ at SM, encoder $\psib$ (mass-only),
         power-counting tiers $T_1,\dots,T_K$, completion floor $\tau$, patience $n_{\rm wait}$
\State pool $P \gets$ working points exciting $T_1$;\quad history $H\gets\emptyset$;\quad $k\gets 1$
\Repeat
  \State $\bm{c}^\star \gets \arg\max_{\bm c\in P}\ \|r_{\bm c}\|\,|\bm{u}_1^\top r_{\bm c}|$
         \Comment{AL completes (or random)}
  \State $H \gets H\cup\{\log w_{\bm c^\star}\}$;\quad $P\gets P\setminus\{\bm c^\star\}$
  \State $\bm{R}\gets[\,r_{\bm c}\ \text{for}\ \log w_{\bm c}\in H\,]$;\quad $(\bm U,\bm S)\gets\mathrm{SVD}(\bm R)$;\quad $\sigma_1\gets S_1$
  \While{$\sigma_1>\tau$} \Comment{physics expands: append the missing direction}
     \State $\psib \gets [\,\psib \mid \bm{u}_1\,]$;\quad recompute $\bm R$ against the new $\psib$;\quad $\sigma_1\gets S_1$
  \EndWhile
  \If{no extension for $n_{\rm wait}$ cycles \textbf{and} $k<K$}
     \State $k\gets k+1$;\quad $P\gets P\cup$ working points exciting $T_k$ \Comment{turn on operators}
  \EndIf
  \State emit Phoenix span $(\sigma_1^{\rm pre},\sigma_1^{\rm post},\sigma_1/\sigma_2,\ \text{fired},\ \text{decision},\ D_\psi)$
\Until{all tiers unlocked and $\sigma_1\le\tau$ is stable}
\end{algorithmic}
\end{algorithm}

The acquisition is in theory space (which $\bm{c}$ to simulate), not on a
kinematic axis. Because the oracle is quadratic, the per-event log-ratios are
precomputed once from a morphing basis, so each candidate evaluation is
closed-form and the full loop runs without re-querying the simulator inside the
inner loop.

\section{Results: tracing the improvement, and monitoring it}
\label{sec:results}

\begin{figure}[t]
\centering
\includegraphics[width=0.74\linewidth]{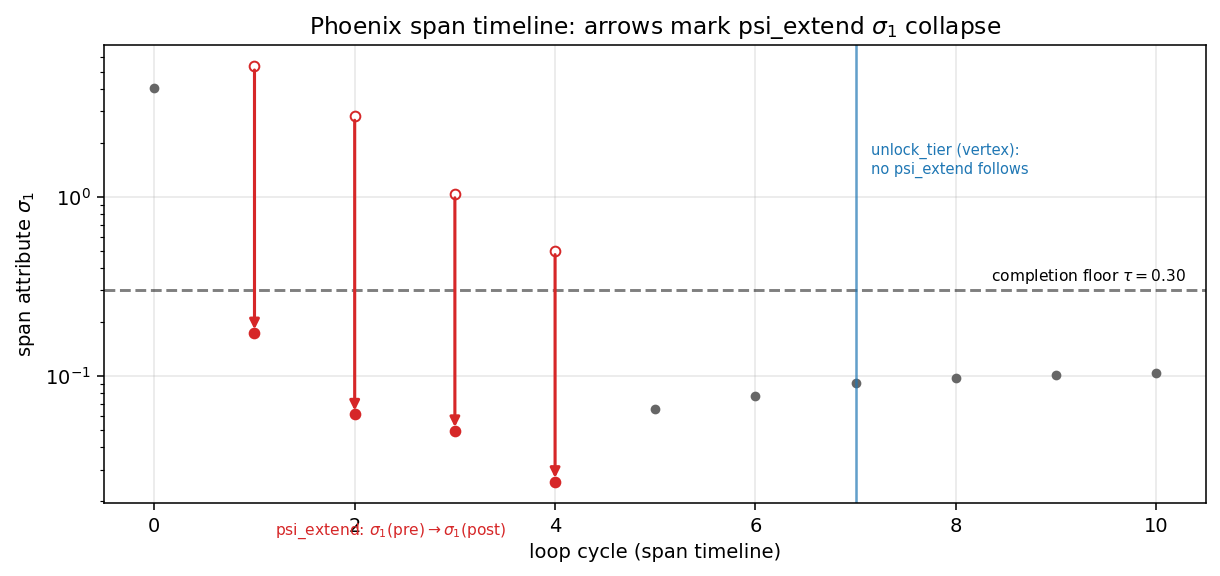}\\[1.5mm]
\includegraphics[width=0.78\linewidth]{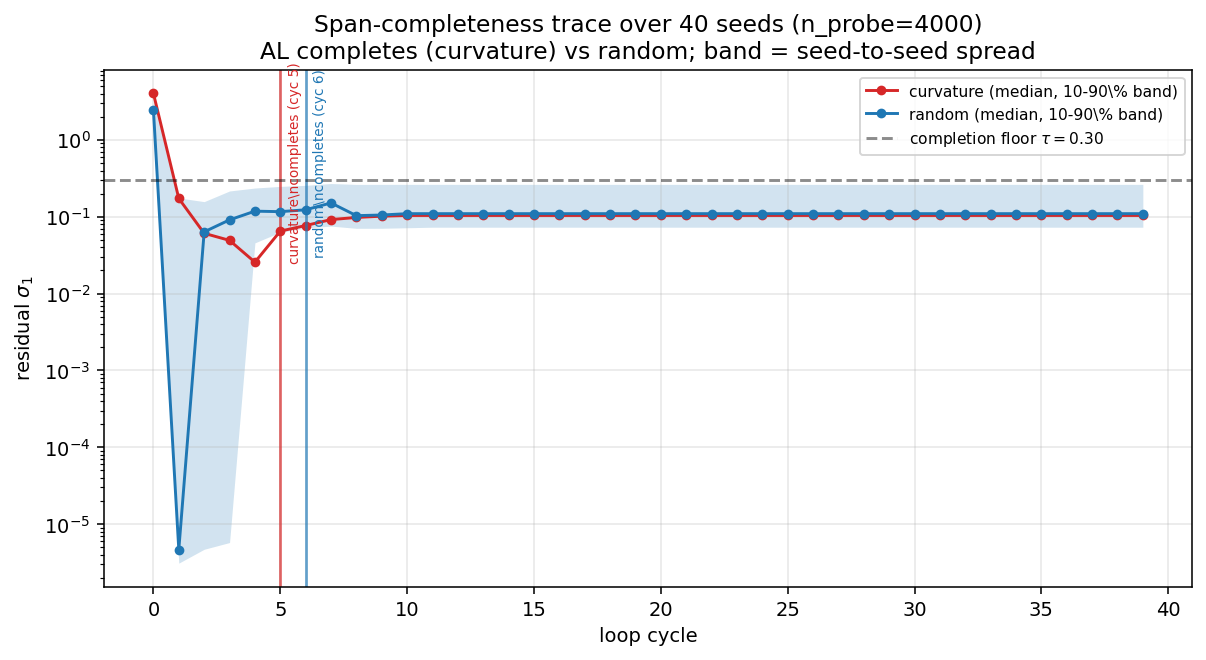}
\caption{\label{fig:monitor} Monitoring the loop. \textbf{(top)} The
Arize-Phoenix span timeline of the residual-energy run, one span per cycle: red
arrows are the four \texttt{psi\_extend} events with their
$\sigma_1(\text{pre})\!\to\!\sigma_1(\text{post})$ collapse (the
learning-correctly read-out); grey dots are cycles with no extension; the
\texttt{unlock\_tier} event at cycle~7 is followed by no further
\texttt{psi\_extend} and a flat sub-floor $\sigma_1$ tail (the learned-completely
read-out). Both proofs are attributes of the monitored spans, not post-hoc
analysis. \textbf{(bottom)} $\sigma_1$ per cycle, median over $40$ seeds with a
$10$--$90\%$ band, residual-energy versus random; vertical lines mark each arm's
completion cycle, the horizontal line the floor $\tau=0.30$. Lower $\sigma_1$ is
not a better arm: the acquisition keeps it elevated to expose the next
out-of-span direction, so the random arm's lower $\sigma_1$ is in-span sampling.
The metric is cycles-to-gate, on which the residual-energy arm is no later than
random (median $5$ versus $6$).}
\end{figure}

The headline of the paper is the recovery of the analytic morphing geometry
(Figure~\ref{fig:recovery}); the residual singular value $\sigma_1$ traced below
is the \emph{completeness diagnostic} that certifies, online, that the recovery
holds across the operator content, not a result in its own right. By the
identifiability proposition the two are one property: $\sigma_1\le\tau$ is the
residual-space reading of the same span-containment the recovery $R^2$ measures.
Table~\ref{tab:trace} and Figure~\ref{fig:monitor} report the loop on the analytic
oracle at $4000$ probe events, started with a mass-only polynomial encoder
($D_\psi=5$), completion floor $\tau=0.30$, both acquisition arms.

\begin{table}[t]
\centering\small
\caption{\label{tab:trace} Residual-energy acquisition arm at $4000$
probe events. Each $\psi$-extension collapses $\sigma_1$; after four extensions
the residual is below floor and stays there. Unlocking the vertex tier at
cycle~7 triggers no further extension, the signature of span-completeness.
$\sigma_1(\text{pre})$ is measured after the cycle's acquisition and before its
$\psi$-extension, $\sigma_1(\text{post})$ after the extension; $\sigma_1(\text{pre})$
may exceed the previous cycle's $\sigma_1(\text{post})$ as the newly acquired
working point exposes structure. Cycles 6 and 8--9 are omitted, with $\sigma_1$
flat below floor across them.}
\begin{tabular}{ccccl}
\toprule
cycle & $D_\psi$ & $\sigma_1$ (pre) & $\sigma_1$ (post) & action \\
\midrule
0  & 5 & 4.03 & 4.03  & fingerprint fires \\
1  & 6 & 5.36 & 0.173 & $\psi$-extend ($31\times$) \\
2  & 7 & 2.81 & 0.061 & $\psi$-extend \\
3  & 8 & 1.04 & 0.049 & $\psi$-extend \\
4  & 9 & 0.50 & 0.026 & $\psi$-extend \\
5  & 9 & 0.065 & 0.065 & complete (watch) \\
7  & 9 & 0.092 & 0.092 & unlock vertex tier; \emph{no extension} \\
10 & 9 & 0.104 & 0.104 & complete \\
\bottomrule
\end{tabular}
\end{table}

Each cycle the acquisition first appends a working point whose out-of-span energy
raises $\sigma_1$ (column ``pre''); the $\psi$-extension then collapses it (column
``post''). $\sigma_1(\text{pre})$ can therefore exceed the previous cycle's
$\sigma_1(\text{post})$: the rise from $4.03$ to $5.36$ between cycles 0 and 1 is
the newly acquired working point exposing structure, not a failure of the
extension. Completion is the cycle after which newly acquired working points no
longer raise $\sigma_1$ above the floor.

\paragraph{Correct learning.}
Each $\psi$-extension appends the dominant residual direction $\bm{u}_1$ and the
leading singular value of the residual operator collapses immediately: $5.36\to
0.173$ at the first extension (a factor $31$), and $4.03\to0.026$ cumulatively
over the four extensions (a factor $\sim\!150$). A collapse of this size is only
possible if $\bm{u}_1$ is the genuine missing direction: appending a wrong or
redundant direction leaves $\sigma_1$ essentially unchanged. The monotone
collapse of $\sigma_1(\text{post})$ across the four extensions
($0.173 \to 0.061 \to 0.049 \to 0.026$) is the correctness certificate: the
model is adding exactly the structure the residual demanded.

\paragraph{Complete learning.}
Completeness is three coincident facts on the monitored trace. (i) $\sigma_1$
falls below the completion floor and \emph{stays} there as new leading-tier
working points are acquired (cycles $5$--$6$): no new four-fermion working point
exposes out-of-span structure. (ii) The fingerprint $\sigma_1/\sigma_2$ stops
firing. (iii) Unlocking the subleading vertex operators at cycle~7, adding
working points that excite $\mathcal{O}_{Hq}^{(1,3)}$, triggers \emph{no} further
extension. Because the vertex operators modify the chiral $Z$-quark couplings and
therefore carry forward-backward-asymmetry structure in $\cos\theta^\star$, which
the encoder's input sees, the absence of a $\psi$-extension on unlock is the
stronger statement that the basis assembled for the four-fermion tier already
spans the vertex direction's angular structure, not merely that the rate shift is
small. The manifold is span-complete with respect to the full operator content,
read directly off the trace rather than asserted.
The acquisition arm reaches this state no later than the random arm
(Table~\ref{tab:trace}, Figure~\ref{fig:monitor}); the run passes all four gates
(extension occurred, all tiers unlocked, manifold completed, acquisition no
slower than random) in $320$\,s.

\paragraph{Monitoring with Phoenix.}
The loop is instrumented as OpenTelemetry/Arize-Phoenix spans (Figure~\ref{fig:loop}).
Each cycle is a span carrying $(\sigma_1^{\rm pre}, \sigma_1^{\rm post},
\sigma_1/\sigma_2, \text{fired}, \text{decision}, D_\psi)$; a $\psi$-extension and
a tier unlock are span events on that cycle. The two proofs are therefore live
read-outs, not post-hoc analyses: \emph{learning correctly} is the
$\sigma_1^{\rm pre}\!\to\!\sigma_1^{\rm post}$ drop attached to every
\texttt{psi\_extend} event, and \emph{learned completely} is the flat
sub-floor $\sigma_1$ timeline together with an \texttt{unlock\_tier} event that
is not followed by any \texttt{psi\_extend}. The controller's decision each cycle
(acquire / extend / unlock) is the span attribute that drives the loop, so the
trace is simultaneously the control signal and the audit log. Figure~\ref{fig:monitor}
renders the span timeline of the run.

\section{Discussion}
\label{sec:discussion}

The separation of roles is what makes the loop work. Active learning is suited to
\emph{completion}, pinning a fixed model's coefficients, and unsuited to
\emph{expansion}, the model-selection question of which new degree of freedom
matters, which a sampling rule cannot answer when the missing direction is weakly
identified. On this analytic oracle the per-event log-ratios are precomputed and
no simulator call is spent inside the loop, so a completion advantage over random
is not expected to appear here; it is expected at higher operator dimension and
with an expensive oracle, where working-point selection costs simulator
evaluations. The contribution is the separation of roles and the completeness
diagnostic, not a sampling speedup. Delegating expansion to the residual
structure, ordered by the SMEFT power counting, removes that demand from the
acquisition entirely. The foundation model's role is correspondingly clarified:
it supplies the representation whose completeness the loop monitors, and its
quality (the $R^2=0.99999$ tangent recovery) is what guarantees the residual
operator measures genuine missing physics rather than encoder error. The
construction is general: any setting with an exact generative structure (here the
SMEFT morphing polynomial) that orders model complexity can drive the same
complete-then-expand loop.

\section{Conclusion}
\label{sec:conclusions}

We presented a foundation model that completes its own representation under an
active-learning loop in which acquisition pins coefficients and the residual
structure, ordered by SMEFT power counting, expands the operator content. On the
analytic Drell-Yan oracle the loop drives the residual singular value down two
orders of magnitude over four $\psi$-extensions, reaches span-completeness, and
demonstrates completeness by the silence of the next operator tier, all read
off a Phoenix-monitored trace that doubles as the controller. The reproduction
code is \texttt{\href{https://github.com/vincecr0ft/ALETHEIA}{https://github.com/vincecr0ft/ALETHEIA}}.
\newpage
\bibliographystyle{unsrtnat}
\bibliography{alethia}

\end{document}